\pdfoutput=1
\documentclass[fleqn]{2017SCGE}
\usepackage[linkcolor=blue,citecolor=blue]{hyperref}

\begin{document}

	\ensubject{subject}

	\ArticleType{Article}
	\Year{2020}
	\Month{February}
	\Vol{60}
	\No{1}
	\DOI{10.1007/xxx}
	\ArtNo{000000}
	
	\title{Microwave diagnostics of magnetic field strengths in solar flaring loops}{Microwave diagnostics of magnetic field strengths in solar flaring loops}
	
	\author[1]{ZHU Rui}{}%
	\author[2,3]{TAN BaoLin}{}
	\author[4,5]{SU YingNa}{}
	\author[1,2]{TIAN Hui}{huitian@pku.edu.cn}
	\author[6]{XU Yu}{}
	\author[2]{\\CHEN XingYao}{}
	\author[2]{SONG YongLiang}{}
    \author[1]{TAN GuangYu}{}
	

	\AuthorMark{Zhu R}
	
	\AuthorCitation{Zhu R, Tan B L, Su Y N, et al.}
	
	\address[1]{School of Earth and Space Sciences, Peking University, Beijing 100871, China; huitian@pku.edu.cn}
	\address[2]{Key Laboratory of Solar Activity, National Astronomical Observatories, Chinese Academy of Sciences, Beijing 100012, China}
	\address[3]{School of Astronomy and Space Science, University of Chinese Academy of Sciences, Beijing 100049, China}
	\address[4]{Key Laboratory of Dark Matter and Space Astronomy, Purple Mountain Observatory, Chinese Academy of Sciences, Nanjing 210008, China}
	\address[5]{School of Astronomy and Space Science, University of Science and Technology of China, Hefei, Anhui 230026, China}
	\address[6]{School of Astronomy and Space Science, Nanjing University, Nanjing 210023, China}

	
	\abstract{We have performed microwave diagnostics of the magnetic field strengths in solar flare loops based on the theory of gyrosynchrotron emission. From Nobeyama Radioheliograph  observations of three flare events at 17 and 34 GHz, we obtained the degree of circular polarization and the spectral index of microwave flux density, which were then used to map the magnetic field strengths in post-flare loops. Our results show that the magnetic field strength typically decreases from $\sim$800 G near the loop footpoints to $\sim$100 G at a height of 10--25 Mm. Comparison of our results with magnetic field modeling using a flux rope insertion method is also discussed. Our study demonstrates the potential of microwave imaging observations, even at only two frequencies, in diagnosing the coronal magnetic field of flaring regions.}
	
	\keywords{solar magnetic field, solar flare, microwave observation, gyrosynchrotron emission}
	

	\maketitle
	

	\begin{multicols}{2}
		\section{Introduction}\label{section1}
		The ubiquitous magnetic field in the solar atmosphere plays a crucial role in various types of physical processes, from small-scale eruptions such as microflares and mini-filament eruptions to large-scale eruptions such as two-ribbon flares and coronal mass ejections (CMEs). All these solar activities are driven by the evolution of the magnetic field. Hence, measurements of the solar magnetic field are very important. However, only the photospheric magnetic field can be measured through the Zeeman effect on a daily basis. The coronal magnetic field has not been routinely measured up to now, due to the large broadening of coronal emission lines and small Zeeman splitting caused by the weaker field. Despite these difficulties, Lin et al. \cite{ref1, ref2} managed to measure the coronal magnetic field in active regions using the infrared coronal emission line Fe XIII 1074.7 nm. However, this method requires a long integration time (order of one hour), and is mainly applicable to the off-limb region of the corona. Coronal magnetic field strengths may also be inferred from observations of oscillations and waves \cite{ref27, ref28, ref29}. However, such measurements can only be used to estimate the average field strength of an individual oscillating structure.
		
		Extrapolation from the measured photospheric magnetic field is a common way to obtain information about the coronal magnetic field. Magnetic field extrapolation methods mainly include the potential field source-surface (PFSS) model \cite{ref3}, linear force-free field (LFFF) models \cite{ref4} and non-linear force-free (NLFFF) models \cite{ref5}. However, one big disadvantage of these magnetic field extrapolation methods is that the force-free assumption is not always valid, especially for the lower atmosphere and flaring regions. Moreover, it is not appropriate to carry out extrapolations for near-limb events since the near-limb measurement of photospheric field is often not reliable.
		
		Attempts have been made to diagnose the coronal magnetic field strengths through radio spectral observations. For instance, assuming that the upper and lower bands of a type II radio burst are produced in the downstream and upstream of a CME shock, respectively, coronal magnetic field strengths could be obtained \cite{ref6, ref7, ref8, ref9}. These results are in good agreement with each other and with some coronal magnetic field models, although the upstream-downstream scenario has been doubted by some reseachers \cite{ref10, ref11}. Observations of many fine structures in radio dynamic spectra, such as type III pair bursts, fiber bursts and Zebra patterns, can also be used to estimate coronal magnetic field strengths \cite{ref12, ref13, ref14}. However, since these observations lack spatial resolution, we do not know which structures the obtained magnetic field strengths correspond to. Moreover, the results depend on the chosen formation mechanisms of the radio fine structures.
		
		Coronal magnetic field strengths can also be derived through microwave imaging observations. In the past few decades, the polarization and spectra of gyroresonance emission were widely used to obtain information of the coronal magnetic field in quiet active regions (ARs) \cite{ref15, ref16, ref17, ref18, ref19}. These ARs usually have strong magnetic fields (typically higher than 100 G), allowing the dominance of gyroresonance emission. Recently, Anfinogentov et al. \cite{ref20} observed a record-breaking magnetic field of $\sim$4000 G at the base of the corona in the unusual active region 12673. Their diagnostics is based on the assumption that the 34 GHz emission is dominated by gyroresonance emission, which was further validated by their simulated microwave images. In weak-field quiet regions, coronal magnetic field strengths could be inferred based on the free-free emission theory at microwave wavelengths \cite{ref21, ref22,ref57}. However, the derived field strengths are larger than those from extrapolations and models, which may be caused by the inaccuracy of the derived polarization degrees. Some other studies are aimed at diagnosing the coronal magnetic field from the infrequently occurring polarization reversal caused by the quasi-transverse propagation effect \cite{ref23, ref24}. These studies have greatly improved our knowledge of the coronal magnetic field in quiet active regions.
		
		However, much less is known for the magnetic field strengths in flaring regions. With the brightness temperature, polarization degree, flux density spectral index and peak frequency obtained from observations of the Nobeyama Radioheliograph (NoRH) and Nobeyama Radio Polarimeters (NoRP), Huang et al. \cite{ref25, ref58, ref59} solved several equations of gyrosynchrotron emission theory to obtain the magnetic field strength, viewing angle and column density of nonthermal electrons. Sharykin et al. \cite{ref60} found inversion of the Stokes-V signal from the NoRH 17 GHz observation of an M1.7 flare. Such a polarization pattern was found to be compatible with the scenario of gyrosynchrotron emission from nonthermal electrons distributed along a twisted magnetic structure, which was revealed from a nonlinear force-free magnetic field (NLFFF) extrapolation. Gary et al. \cite{ref26} reported the first observation of microwave imaging spectroscopy from the Expanded Owens Valley Solar Array (EOVSA), which has a high spatial (6$''$--25.7$''$), spectral (160 MHz), and temporal resolution (1s) over a broad frequency range (3.4--18 GHz). Interpreting the microwave emission as gyrosynchrotron emission, they obtained the magnetic field strengths at four pixels in the region of an X-class flare by fitting the observed flux density spectra. The fitting results depend on the initial values of various coronal parameters and the image quality (signal-to-noise ratio) to some degree, but they indeed obtained reasonable results. The fitting enables them to obtain the spectral indices of nonthermal electrons, which are consistent with the results inferred from observations of the Reuven Ramaty High Energy Solar Spectroscopic Imager (RHESSI). Note that currently EOVSA is unable to give the polarization information of the microwave emission. The magnetic field strengths of flare loops have also been measured using other methods. For example, Kuridze et al. \cite{ref47} obtained a magnetic field strength of about 350 G at a height of 25 Mm using high-resolution imaging spectropolarimetry from the Swedish 1 m Solar Telescope. Li et al. \cite{ref48} estimated the magnetic field strength of a flare loop to be about 120 G at a height of 35 Mm using the magnetohydrodynamic seismology technique.
		
		In this paper, we present results from the microwave diagnostics of magnetic field strengths in several post-flare loops based on NoRH observations at 17 and 34 GHz, using the formula of polarization degree derived by Dulk \cite{ref30}. Section 2 briefly introduces the instruments. Section 3 describes our method and results. The summary and discussion are given in Section 4.
		
		
		
		\section{Instruments}\label{sec:2}
		NoRH, a Sun-dedicated radio telescope, provides microwave images of the full-disk Sun at two frequencies 17 GHz and 34 GHz, with a spatial resolution of $\sim$10$''$ and $\sim$5$''$, respectively. The cadence is 1 s for normal observations and can reach 0.1 s for event observations. The radioheliogragh consists of 84 parabolic antennas with an 80-cm diameter \cite{ref31} and usually observes the Sun from 22:50 UT to 06:20 UT on the next day. Intensity images (Stokes I) can be obtained at both frequencies, and only the 17 GHz observations contain polarization information (Stokes V). 
		
		We also used data from the Atmospheric Imaging Assembly (AIA; Lemen et al. \cite{ref32}) and the Helioseismic and Magnetic Imager (HMI; Scherrer et al. \cite{ref33}) on board the Solar Dynamics Observatory (SDO; Pesnell \cite{ref34}), the Transition Region and Coronal Explorer (TRACE; Handy et al. \cite{ref35}) and the Michelson Doppler Imager (MDI; Scherrer
		et al. \cite{ref36}) on board the Solar and Heliospheric Observatory (SOHO; Domingo et al. \cite{ref37}) . These instruments provide full-disk UV and EUV images and photospheric line-of-sight (LOS) magnetograms, complementary to the microwave data.
		
		As a common practice \cite{ref61, ref62}, we aligned the NoRH and AIA/TRACE images using the coordinate information in the headers of the image FITS files. A comparison between the flare loops in the NoRH images and EUV images (see the next Section) confirms the accuracy of the alignment.

		
		\section{Method and results}\label{sec:3}
		\subsection{Method}
		Gyrosynchrotron emission is produced by mildly relativistic electrons (Lorentz factor $ \gamma < 3$) due to their gyration around magnetic field lines. In the standard picture of solar flares, also known as the CSHKP model \cite{ref38, ref39, ref40, ref41}, electrons are accelerated at or near the magnetic reconnection site and propagate downward along the magnetic field lines into the dense chromosphere, losing most of their energy there and heating the chromosphere, which results in chromospheric evaporation and formation of hot EUV/soft X-ray flare loops. Considering the fact that the nonthermal electrons exist in a region where the magnetic field is not too weak (at least tens of Gauss), the mechanism of microwave emission would be gyrosynchrotron. This is usually the case during the flare impulsive phase. For the later gradual phase, nonthermal electrons could be thermalized due to Coulomb collisions with ambient particles, and hence the free-free emission may dominate \cite{ref42, ref43}. We can distinguish between thermal and nonthermal microwave emission by examining the spectral index ($ \alpha $) of microwave flux density (assuming a power-law distribution of flux density): if the emission mechanism is free-free, $ \alpha $ is about zero or slightly larger or less than zero; if the emission mechanism is gyrosynchrotron, $ \alpha $ is negative with a typical value of -2 \cite{ref30}. 
		
		If the observed microwave emission is dominated by nonthermal gyrosynchrotron emission, the following equation can be used to estimate the magnetic field strength \cite{ref30}:
		
		\begin{equation}
		\textmd{r}_{\rm c}\approx1.26\times10^{0.035\delta}\times10^{-0.071\cos \theta}\;(\frac{\nu}{\nu_B})^{-0.782+0.545\cos\theta}\,
		\label{eq:1}
		\end{equation}
		
		where $ \textmd{r}_{\rm c} $ is the circular polarization degree of the microwave emission, $ \delta $ is the power-law spectral index of nonthermal electrons, $ \theta $ is the viewing angle between the LOS and the direction of magnetic field, $ \nu $ is the observation frequency (17 GHz in this paper) and $ \nu_B $ is the electron gyrofrequency ($ \nu_B = eB/2\pi m_e c \approx 2.8B$ MHz, $B$ is the magnetic field strength in the unit of Gauss). Here, $ \delta $ has the following relationship with $ \alpha$:
		
		\begin{equation}
		\delta = -1.1(\alpha-1.23)
		\label{eq:2}
		\end{equation}
		
		Except the magnetic field strength, the only uncertain parameter in \cref{eq:1} is the viewing angle $ \theta $, which we have tried to determine from magnetic field modeling using the flux rope insertion method developed by van Ballegooijen \cite{ref63}. We first computed the potential fields from SDO/HMI magnetic field observations, and then inserted thin flux bundles along selected filament paths or polarity inversion lines. The next step was to use magneto-frictional relaxation to drive the field toward a force-free state. We have constructed a series of magnetic field models by adjusting the axial and poloidal flux until a good match of the model field lines with the observed flare loops is achieved. For a more detailed description of this method, we refer to the recent publications of Su et al. \cite{ref64} and Chen et al. \cite{ref65}. Using the best-fit model we could have a reasonable estimation of the viewing angle for each flare loop. 
		
		It should be noted that \cref{eq:1} is applicable only for $ 10\leq s \leq 100 $ and $ \tau \ll 1$, where $ s=\nu/\nu_B $ is the harmonic number and $ \tau$ is the optical depth. In our cases, the observed NoRH brightness temperature is obviously less than the actual coronal temperature, which is several MK for quiet regions and about 10 MK for flaring regions, suggesting that the condition of being optically thin ($ \tau \ll 1$) is often satisfied. However, if the magnetic field strength is too small, the harmonic number could be larger than 100. Therefore, after calculating the magnetic field strengths using \cref{eq:1}, we should remove pixels where $s$ is larger than 100.
		
		\subsection{An M5.8 flare on March 10, 2015}

\begin{figure}[H]
			\includegraphics[scale=0.395]{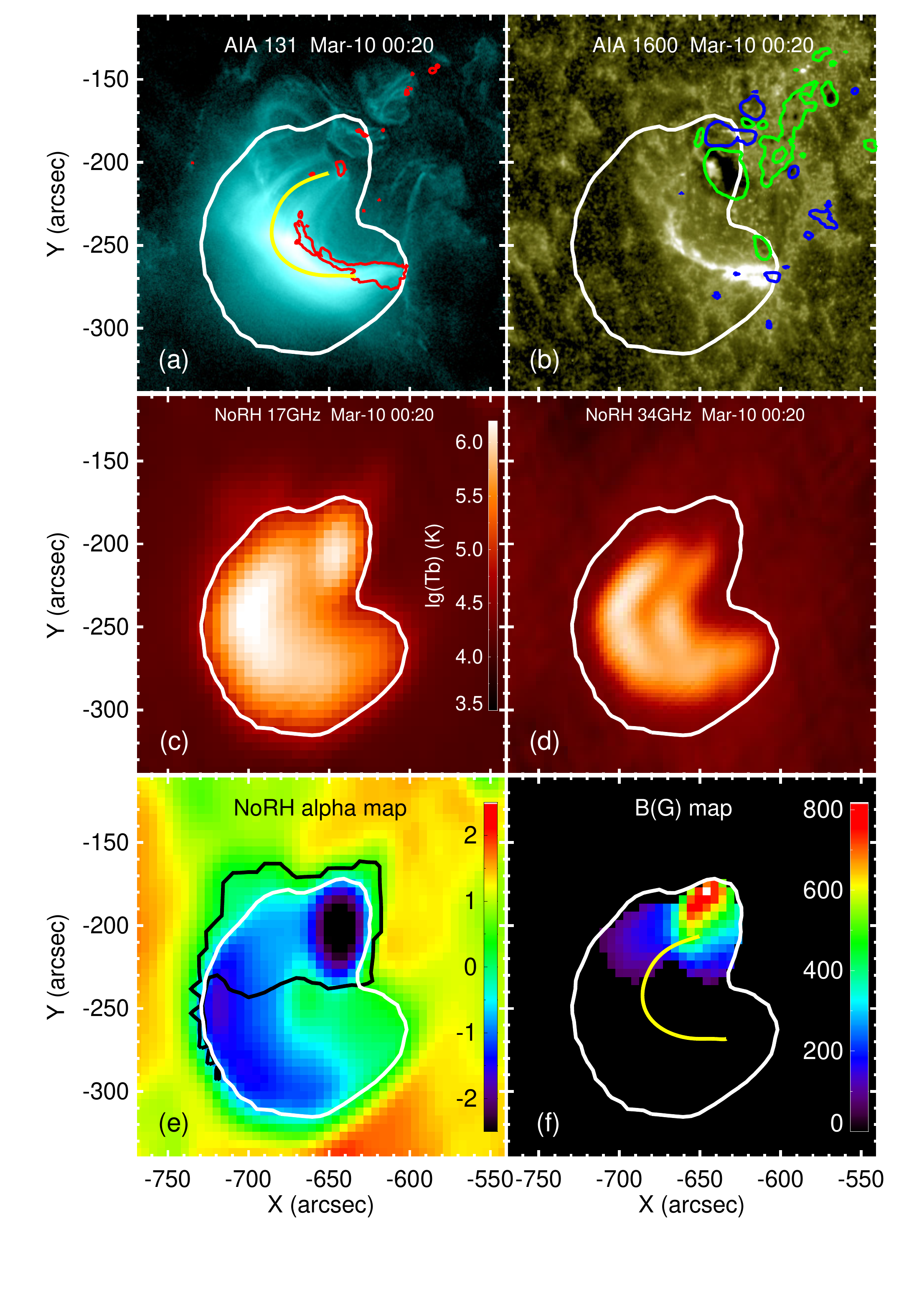}
			\caption{(a)-(d) SDO/AIA 131 $\AA$ and 1600 $\AA$, NoRH 17 GHz and 34 GHz images of the event at 00:20 UT on March 10, 2015. The white curve in each panel represents the contour of the NoRH 17 GHz brightness temperature. The red contours indicate the brightest region in the 1600 $\AA$ image, and the green (blue) contours mark strong positive (negative) photospheric longitudinal magnetic flux density at the 300 (-300) G level observed by SDO/HMI. (e) Spatial distribution of $\alpha$. The black contour indicates the level of 5$\%$ circular polarization degree. (f) Map of magnetic field strength. The loop-like yellow curve in (a) and (f), which is used for  viewing angle estimation and indicates a track along which the variation of magnetic field strength is plotted in \cref{fig:example4}, is one of the field lines from our best-fit magnetic field model.} 
			\label{fig:example1}
		\end{figure}

		An M5.8 flare occurred in NOAA active region (AR) 12297 (position: S17E39) late on March 9, 2015 and early on the next day. It was well observed in microwave by NoRH and in EUV by SDO/AIA. The flare started from 23:29 UT and peaked at 23:53 UT on March 9, 2015, with a total duration of about 1 hour. \cref{fig:example1}(a) and (b) show the AIA 131 $\AA$ and 1600 $\AA$ images of the event taken at 00:20 UT on 2015 March 10. A bright flare loop is clearly visible in the 131 $\AA$ image. The loop footpoints spatially correspond to the bright ribbons in the 1600 $\AA$ image, though the southern footpoint occupies a much larger area compared to the compact northern one. \cref{fig:example1}(c) and (d) show the NoRH microwave images of the flare loop at 17 and 34 GHz, respectively, at the same time. The white curve in each panel is the contour of the 17 GHz brightness temperature at the level of lg$(T_B)=4.8$, which outlines the approximate shape of the flare loop in microwave. The maximum brightness temperature at 17 GHz is about 1-2 MK, suggesting that the emission is optically thin. Although the flare had evolved into the decay phase at this moment, the flare loop can be clearly observed with a well-defined morphology. It should be noted that the microwave flare loop at 17 GHz (\cref{fig:example1}(c)) occupies a larger area than its EUV counterpart (\cref{fig:example1}(a)) and that the northern footpoint is very bright in microwave but not in EUV, which will be discussed in the next section.
		
		
		Since we have microwave observations at both 17 GHz and 34 GHz, we can calculate the spectral index $\alpha$ by assuming a power-law distribution of the flux density ($F\propto \nu^{-\alpha}$). The spatial distribution of the spectral index is shown in \cref{fig:example1}(e). It is obvious that $\alpha$ is negative at the looptop and the northern footpoint, meaning that the microwave emission is dominated by nonthermal gyrosynchrotron emission. The value of $\alpha$ at the southern footpoint is around zero, indicating the thermal nature of the microwave emission there. Because the observational accuracy of circular polarization measured by NoRH is $\sim$5$\%$, our diagnostics is limited to the region with a circular polarization degree higher than 5$\%$, as outlined by the black contour in \cref{fig:example1}(e).

		Looking back to \cref{eq:1}, we can find that, once $ \delta $ is derived from the observation and $ \theta $ is determined from the magnetic field modeling, lower polarization degrees correspond to weaker magnetic field strengths and higher harmonic numbers. Therefore, if the polarization degree is not high enough, i.e., the harmonic number might be larger than 100, then \cref{eq:1} is no longer applicable. In other words, we can only estimate the magnetic field strengths in regions where the polarization degree is high enough, i.e., the area within the black contour in \cref{fig:example1}(e). The loop-like yellow curve shown in \cref{fig:example1}(a) and (f) is one of the magnetic field lines from our best-fit magnetic field model. It is used to estimate the viewing angle. \cref{fig:example1}(f) shows the diagnostic result by using the average viewing angle ($\theta$ $\sim75^\circ$) of the section of this field line within the black contour. The magnetic field strength appears to reach $\sim$800 G near the northern footpoint and decrease with height along the loop. Due to the low polarization degree at the looptop and the southern footpoint, we cannot diagnose the magnetic fields there. Note that hereafter we use the colorbar of rainbow plus white to plot the map of magnetic field strength, and the white pixel marks the maximum field strength. 
		
		\subsection{An M1.7 flare on July 10, 2012}

\begin{figure}[H]
			\centering
			\includegraphics[scale=0.395]{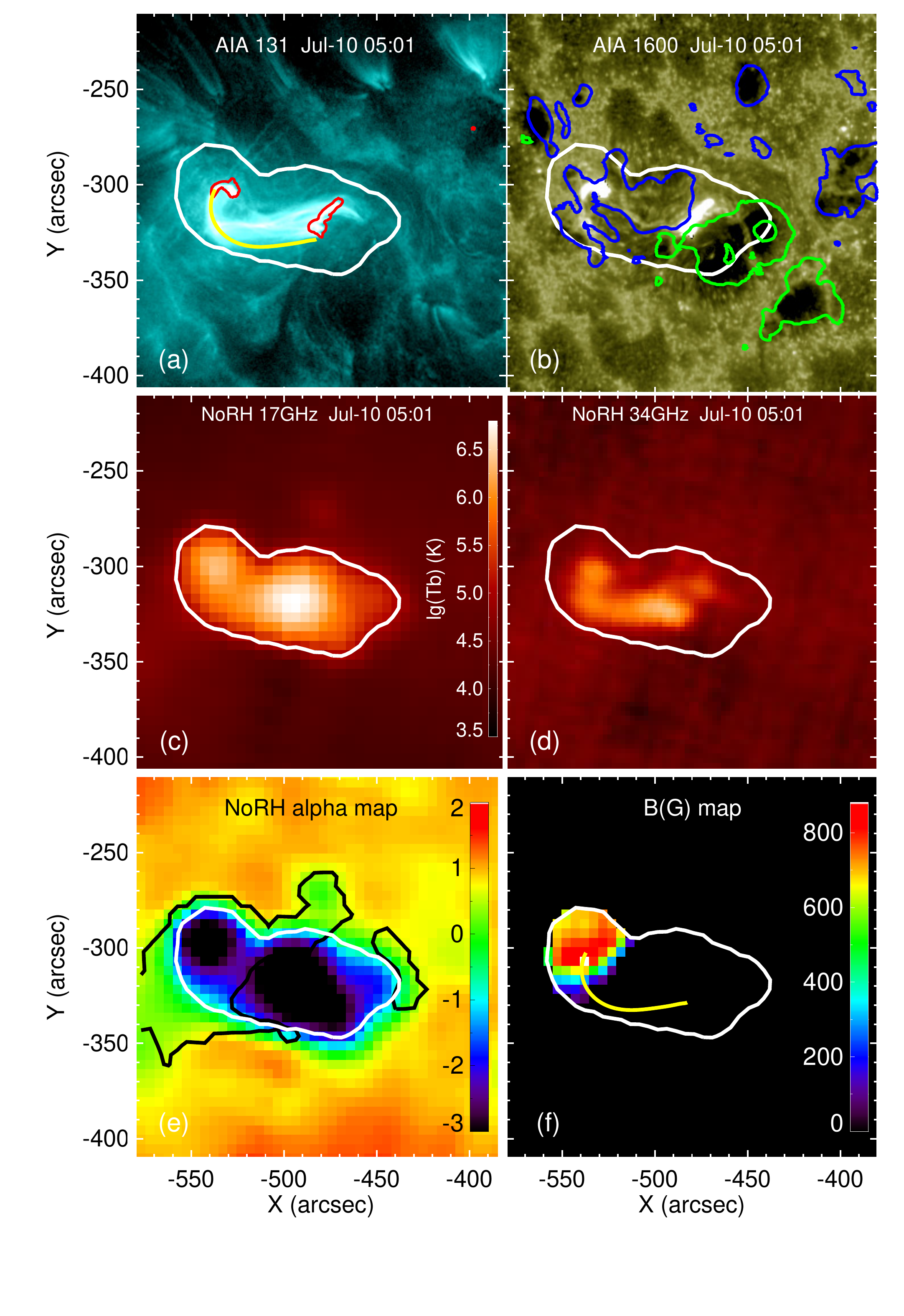}
			\caption{Similar to \cref{fig:example1} but for the event observed around 05:01 UT on July 10, 2012.} 
			\label{fig:example2}
		\end{figure}

		An M1.7 flare peaked around 04:58 UT on July 10, 2012 in NOAA active region 11520 (S17E19). It was also well observed in microwave by NoRH and in EUV by SDO/AIA. \cref{fig:example2}(a) and (b) show the AIA 131 $\AA$ image and 1600 $\AA$ images of the event at 05:01 UT, respectively. A bright flare loop is clearly visible in the 131 $\AA$ image, and the loop footpoints spatially correspond to the bright flare ribbons seen in the 1600 $\AA$ image. \cref{fig:example2}(c) and (d) show the NoRH microwave images of the flare loop at 17 and 34 GHz at 05:01 UT, respectively. The white curve in each panel is the contour of the 17 GHz brightness temperature at the level of lg$(T_B)=5.0$, which outlines the approximate shape of the flare loop in microwave. Interestingly, the flare loop as seen in the AIA 131 $\AA$ image (\cref{fig:example2}(c)) looks very similar to the microwave flare loop at 34 GHz (\cref{fig:example2}(d)). The microwave flare loop at 17 GHz (\cref{fig:example2}(c)) occupies a larger area than its EUV counterpart (\cref{fig:example2}(a)), which is likely due to the lower spatial resolution of the 17 GHz image. The maximum brightness temperature at 17 GHz is about 3-4 MK, suggesting that the emission is optically thin.
		
		 \noindent 
	
		Similarly, we then calculated the spectral index $\alpha$, as shown in \cref{fig:example2}(e). It is clear that $\alpha$ is negative for almost the entire flare loop, meaning that the microwave emission is dominated by nonthermal gyrosynchrotron emission. The black contours in \cref{fig:example2}(e) indicate the 5$\%$ level of the circular polarization degree. It is interesting that the polarization degree is high near the eastern footpoint and low near the western one, which will be discussed in the next section. The flare loop reveals a sigmoid structure, which appears to consist of two J-shaped loops. The loop-like yellow curve shown in \cref{fig:example2}(a) and (f) is one of the magnetic field lines from the best-fit magnetic field model. Again, here we used a fixed viewing angle ($\theta$ $\sim$ $61^\circ$), which is the mean value of the viewing angles along the section of this field line within the black contour. Using \cref{eq:1}, we obtained the map of magnetic field strength, as shown in \cref{fig:example2}(f). The magnetic field strength is about 850 G near the eastern footpoint and decreases with height along the loop leg. Due to the low polarization degree near the western footpoint, we cannot diagnose the magnetic field there. 

		\subsection{An M4.4 flare on September 16, 2005}
		
			\begin{figure}[H]
			\centering
			\includegraphics[scale=0.39]{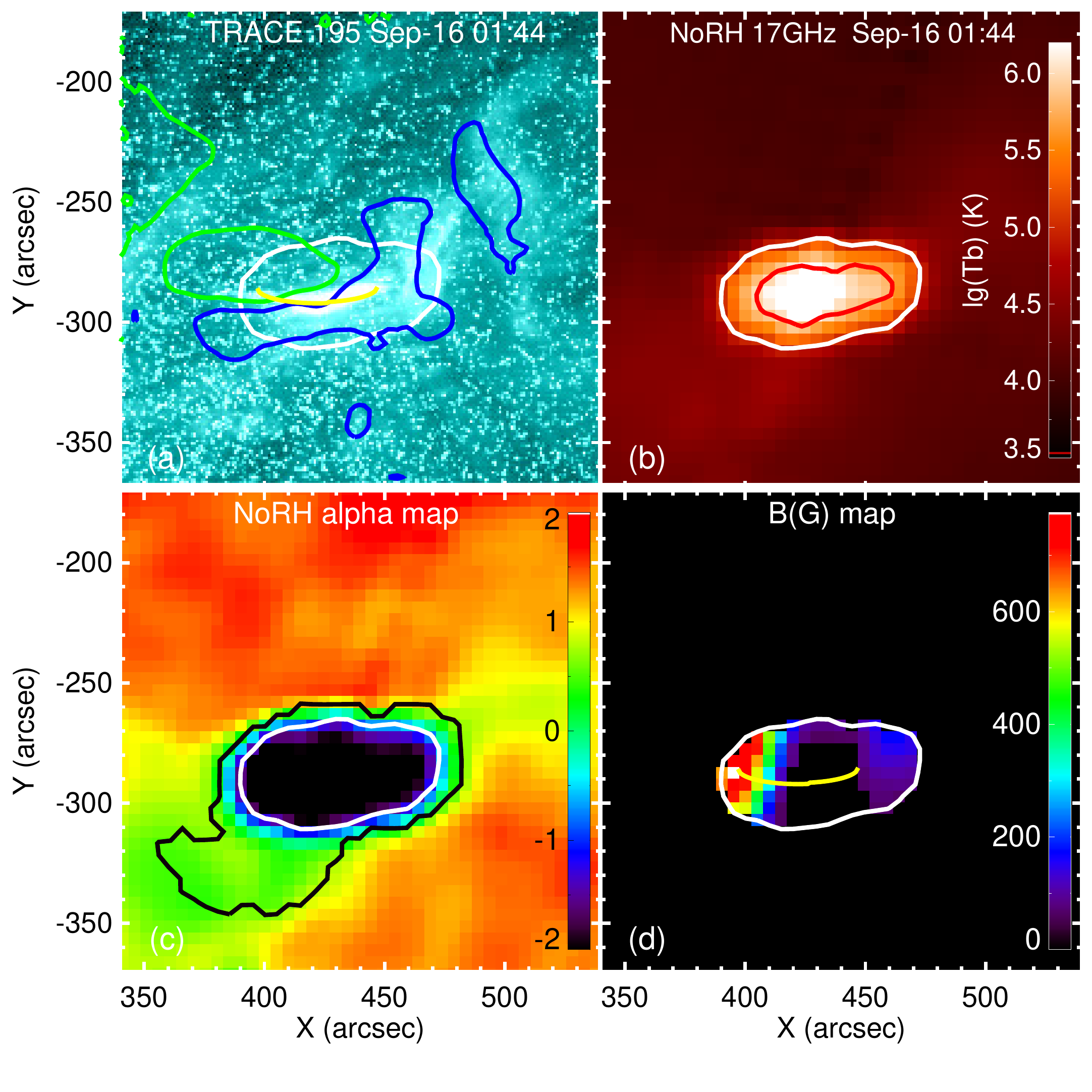}
			\caption{(a)-(b) TRACE 195 $\AA$ and NoRH 17 GHz image at 01:44 UT on September 16, 2005. The green (blue) contours outline strong positive (negative) photospheric longitudinal magnetic flux density \textbf{at the 300 (-300) G level} observed by SOHO/MDI at 00:03 UT. The red contour outlines the region of high brightness temperature at NoRH 34 GHz. (c)-(d) Similar to \cref{fig:example1}(e)-(f). The loop-like yellow curve in (a) and (d), which is used for viewing angle estimation and indicates a track along which the variation of magnetic field strength is plotted in \cref{fig:example4}, marks a semi-circular loop that best fits the observed flare loop in the 195 $\AA$ image.}
			\label{fig:example3}
		\end{figure} \noindent 
		
		An M4.4 flare occurred in NOAA AR 10808 (S11W36) and peaked around 01:47 UT on September 16, 2005. The flare was well observed in microwave by NoRH. Except TRACE 195 $\AA$ images, no other high-resolution (E)UV images are available for this event. \cref{fig:example3}(a) and (b) show the TRACE 195 $\AA$ and NoRH 17 GHz images of the event at 01:44 UT, respectively. A small bright flare loop is visible in the TRACE 195 $\AA$ image. The whole flare loop is manifested as a bright elongated feature in the microwave image. The green (blue) contours outline strong positive (negative) photospheric longitudinal magnetic fields observed by SOHO/MDI about one hour before the flare (00:03 UT). There is no MDI data around 01:44 UT. The red contour overplotted in the 17 GHz image outlines the region of high brightness temperature at NoRH 34 GHz. The 34 GHz contour appears to have two components, likely corresponding to the two footpoints of the flare loop.

		\cref{fig:example3}(c) shows the spatial distribution of $\alpha$. It is obvious that $\alpha$ is negative for the entire flare loop, meaning that the microwave emission is dominated by nonthermal gyrosynchrotron emission. The black curve in \cref{fig:example3}(c) is the 5$\%$ circular polarization degree contour. The white contour outlines the region of 17 GHz brightness temperature higher than lg$(T_B)=5.0$. The maximum brightness temperature at 17 GHz is about 1-2 MK, suggesting that the emission is optically thin. Since the magnetogram was taken $\sim$1.5 h before the flare and the magnetic field may have evolved a lot since then, we decided not to perform the magnetic field modeling. Instead, we assumed that the flare loop is semi-circular and perpendicular to the local surface. The height of the loop was chosen in such a way that the loop projected on the solar surface fits the observed flare loop in the 195 $\AA$ image. The best-fit loop is marked by the yellow curve in \cref{fig:example3}(a) and (d). The viewing angle was taken to be $\sim$ $54^\circ$, which is the mean value of the viewing angles along the loop legs. From  \cref{fig:example3}(d) we can see that the magnetic field strength is about 750 (200) G near the eastern (western) footpoint and it decreases with height along both loop legs. The magnetic field strengths at the loop top can not be obtained since $s$ is larger than 100 there.

		\subsection{Variation of magnetic field strength with height}
		Before discussing the variation of magnetic field strength with height, it is necessary to evaluate the uncertainty of our measurements. One source of uncertainty comes from the possible contribution of thermal emission to the observed radio flux. We may roughly estimate the thermal contribution from observations of the first event, in which the northern loop leg is dominated by nonthermal gyrosynchrotron emission while the southern one by thermal free-free emission. We find that the brightness temperature $T_B$ ($\sim$10$^6$ K) in the northern leg is larger than the $T_B$ ($\sim$10$^5$ K) in the southern one by one order of magnitude at 17 GHz. And the $T_B$ ($\sim$4$\times$10$^5$ K) in the northern leg is also larger than the $T_B$ ($\sim$6$\times$10$^4$ K) in the southern one at 34 GHz. In fact, if the thermal free-free $T_B$ reaches 1 MK at 17 GHz, the emission measure (EM) will be $\sim$10$^{33}$ cm$^{-5}$ according to Dulk \cite{ref30}, much larger than a typical EM of $\sim$10$^{28}$  cm$^{-5}$ during flares. Moreover, thermal gyroresonance emission is also unlikely to dominate in our three cases. According to the gyroresonance frequency formula: $\nu_B=2.8\times10^6 sB$ (Hz), where the harmonic number $s$ is usually 2 or 3, if the 17 GHz emission is dominated by a gyroresonance component, the coronal magnetic field strength would be about 2000-3000 G, which is very unlikely. All these suggest that the observed radio flux in our cases is dominated by nonthermal gyrosynchrotron emission. If we approximate the thermal contribution in the northern leg by the observed radio flux in the southern one in our first case, we can roughly estimate the relative error in the nonthermal gyrosynchrotron emission caused by thermal contribution to be 10$\%$ (15$\%$) at 17 GHz (34 GHz).

		\begin{figure}[H]
			\centering
			\includegraphics[scale=0.39]{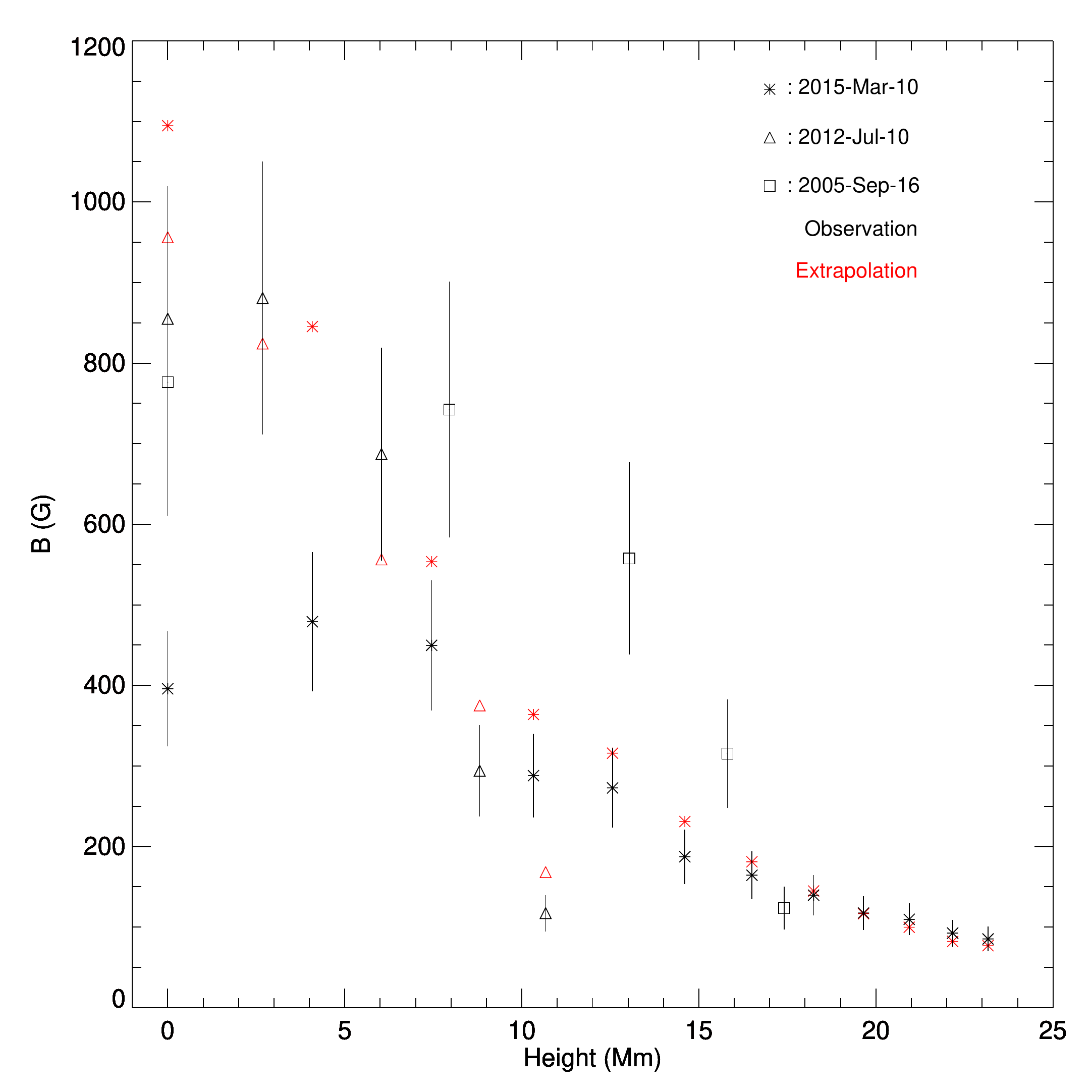}
			\caption{Variation of magnetic field strength with height for the three yellow loops marked in \cref{fig:example1}-\cref{fig:example3}. The asterisks, triangles and squares stand for results from the 2015 March 10, 2012 July 10 and 2005 September 16 observations, respectively. The black (red) symbols indicate the diagnosed (modeled) magnetic field strength.}
			\label{fig:example4}
		\end{figure}
		\noindent

		The instrumental error could be evaluated as the relative fluctuation of brightness temperature (or microwave flux density that is proportional to brightness temperature) in a quiet region. For our three observations, we found a fluctuation level of 0.97$\%$ (2.2$\%$), 0.93$\%$ (3$\%$) and 1.09$\%$ (4.9$\%$), respectively, at 17 GHz (34 GHz). Both the thermal contribution and instrumental error could lead to uncertainty of the spectral index ($\alpha = -ln(Flux_{17}/Flux_{34})/ln2)$. From the error propagation theory, we found that the relative (absolute) error of $\alpha$  is $\sim$15$\%$ ($\sim$0.3) for a typical $\alpha$ value of -2. The error of viewing angle could be reasonably obtained by examining the deviation of the viewing angles in loop legs from the average value used in our calculation. These angles are mostly within the range of $\pm$15$^\circ$ with respect to the mean value in each case. This range was thus taken as the error of the viewing angle. In addition, the polarization degree has an uncertainty of $\sim$5$\%$. By propagating all these errors to the magnetic field strength, we found that the relative error of the derived $B$ is $\sim$18$\%$, $\sim$19$\%$ and $\sim$21$\%$ for the three events, respectively.
		
		For each event presented above, we have shown the spatial distribution of magnetic field strength in part of the flare loop. Due to the low polarization degree in some regions, we cannot obtain the magnetic field strength for the whole flare loop. For each event, we plotted the variation of magnetic field strength along the selected yellow loop. The results are shown in \cref{fig:example4}. We also compared our results with the field strengths in the best-fit magnetic field models for the first two cases and found that they are of the same order of magnitude, especially at high altitudes in the first case. In the first case the field line along which we chose to plot the variation of magnetic field strength does not pass the location of the maximum diagnosed magnetic field, leading to $\sim$1000 G of modeled field and only $\sim$400 G of diagnosed field at lower heights. The spatial offset of the strongest diagnosed field from the strongest photospheric field will be discussed in the next Section. Despite the general agreement between results from our new method and the magnetic field modeling, we caution that a detailed comparison is meaningless because the force-free assumption may not be valid in the flare loops studied here.

		\section{Summary and discussion}\label{sec:4}
		
		Using the polarization equation (\cref{eq:1}) derived from the gyrosynchrotron emission theory \cite{ref30}, we have estimated the magnetic field strengths of three flare loops. Our study demonstrates the potential of microwave imaging observations, event at only two  frequencies, in diagnosing the coronal magnetic field of flaring regions. Since \cref{eq:1} is only applicable to regions with a non-negligible circular polarization degree, we can only derive the magnetic field strengths in partial regions of the flare loops, i.e., one or both legs of the flare loop in each of our three cases. Our results show that the magnetic field strength typically decreases from $\sim$800 G near the loop footpoints to $\sim$100 G at a height of 10--25 Mm.

		As mentioned above, we could not derive the magnetic field strengths in regions with a very low polarization degree. Why do some regions show a low polarization degree even though the emission is dominated by nonthermal gyrosynchrotron emission? This is probably related to the coarse spatial resolution of our microwave observations. In other words, the observed brightness temperature at each pixel is a kind of averaged value over a larger area. Due to such averaging effect, even if there are sub-resolution high-polarization microwave sources, low polarization degree is still likely to be observed. Anther factor is the propagation effect. When there is quasi-transverse field in the corona through which the microwave emission propagates, the polarization degree could be changed \cite{ref23, ref24}. 

		In the first event, the strongest magnetic field derived from the microwave observations appears to be not cospatial with the northern footpoint of the EUV flare loop (\cref{fig:example1}). Two reasons may account for such a spatial offset. First, NoRH observes structures where nonthermal electrons are concentrated, whereas the AIA filters sample mainly the thermalized plasma in the flare loop. Second, the photospheric magnetic field configuration at both loop footpoints is complicated, i.e., there are magnetic features with opposite polarities around both footpoints. Possibly, there are two loop systems with different directions of magnetic field, which appear to be visible from the AIA 131 $\AA$ and NoRH 34 GHz images (\cref{fig:example1}(a) and (d)). Since the 17 GHz microwave observation has a low spatial resolution and hence cannot resolve these loops, we may see a deviation of the maximum magnetic field strength from the northern footpoint identified from the (E)UV images. In the other two events, the highest magnetic field strengths derived from the microwave observations are nearly cospatial with the loop footpoints. Also, in these two events, the photospheric magnetic configuration at the two footpoints is simple, i.e., there is only one magnetic polarity at each footpoint.
		
		Another interesting feature in the first event is that the southern leg of the flare loop, which has weak microwave emission, is very bright in EUV. While the northern leg of the flare loop, which has very weak EUV emission, is very bright in microwave. This result can be explained by the different emission mechanisms of EUV and microwave. \cref{fig:example1}(e) shows that the nonthermal microwave emission (negative $\alpha$) dominates the loop segment from the looptop to the northern footpoint, and that the thermal component ($\alpha$ $\approx$ 0) dominates the southern footpoint and part of the southern loop leg. Since the EUV emission is thermal, and nonthermal microwave emission is usually stronger than thermal microwave emission, the brightest microwave and EUV emission comes from the northern and southern legs of the flare loop, respectively. In principle, we could infer the magnetic field strengths near the southern footpoint based on the free-free emission theory. However, the polarization degree is too low ($\sim$1$\%$) to allow a reliable measurement. Another question is: why do the two footpoints have different emission mechanisms? We noticed that the flare has evolved into the decay phase, during which stage nonthermal electrons are gradually thermalized due to Colomb collisions with ambient particles. The two footpoints likely have different physical parameters (magnetic field, density, et al.), meaning that the time scales of thermalization could be highly different between the two footpoints. As a result, the thermalization process may be achieved earlier at the southern footpoint compared to the northern one.
		
		
		Despite the fact that NoRH has been in operation for more than 20 years, there are very limited numbers of observations where the microwave flare loops were imaged. In some cases, NoRH observed flare loops, but the emission appears to be dominated by thermal free-free emission rather than nonthermal gyrosynchrotron emission. Only when nonthermal electrons are trapped by the magnetic field lines in flare loops could the nonthermal microwave emission be observed. Conditions for flare loops to trap nonthermal electrons are still not well understood, but one favoring condition is a large magnetic mirror ratio. In our three cases, the maximum magnetic field strength is $\sim$800 G at the loop footpoint, and the minimum field strength at the looptop might be several G on the basis of the tendency shown in \cref{fig:example4}. Therefore, the magnetic mirror ratios in our cases appear to be very large, allowing nonthermal electrons to be trapped in the flare loops.
		
		
		We noticed that the full capability of radio imaging spectroscopy has been realized with the completion of the Mingantu Spectral Radioheliograph (MUSER; Yan et al. \cite{ref49}), the Low Frequency Array (LOFAR; van Haarlem et al. \cite{ref50}), the Murchison Widefield Array (MWA; Tingay et al. \cite{ref51}) and the upgraded Karl G. Jansky Very Large Array (VLA; Perley et al. \cite{ref52}). Equipped with these new radio telescopes, we are now able to uncover more about various physical processes in the solar atmosphere, including not only flares \cite{ref55} but also small-scale activities such as coronal jets \cite{ref53}, transient coronal brightenings \cite{ref54}, UV bursts \cite{ref56} and solar tornadoes \cite{ref70}. 
		
		\Acknowledgements{This work is supported by the Strategic Priority Research Program of CAS with grant XDA17040507, NSFC grants 11790301, 11790302, 11790304, 11825301, 11973057, 11803002 and 11473071. We thank Dr. Yang Guo for helpful discussion. }

		\bibliographystyle{unsrt}

	\end{multicols}
\end{document}